\documentclass{easychair}

\usepackage[utf8]{inputenc}
\usepackage{textcomp}
\usepackage{xspace}
\usepackage{enumitem}
\usepackage{listings}

\title{The \sh{remote_build} Tool%
\thanks{This work is supported by the Austrian Science Fund (FWF) project P27502.}}

\author{Christian Sternagel}

\institute{
  Department of Computer Science\\
  University of Innsbruck, Innsbruck, Austria\\
  \email{christian.sternagel@uibk.ac.at}}
  
\authorrunning{C.~Sternagel}

\titlerunning{\RB}

\newcommand\tool[1]{\nolinkurl{#1}}
\newcommand\file[1]{\nolinkurl{#1}}
\newcommand\isaforversion{2.32\xspace}
\newcommand\isaversion{Isabelle2017}
\newcommand{\doi}[1]{
  \href{http://dx.doi.org/#1}{\nolinkurl{doi:#1}}}

\newcommand\texttildemid{\raisebox{.553ex}{\hbox{\texttildelow}}}

\lstdefinestyle{bashstyle}{%
  upquote=true,%
  basicstyle=\ttfamily\color{black},%
  basewidth={0.5em,0.45em},%
  literate=*%
    {\~}{{{\color{black}\texttildemid}}}1
}

\newcommand{\sh}[1]{\text{%
  \lstinline[style=bashstyle,%
    basicstyle=\ttfamily\color{black}]^#1^}}

\lstnewenvironment{bash}{%
  \lstset{style=bashstyle,aboveskip=\smallskipamount,belowskip=\medskipamount}%
}{}

\newcommand{\isafor}{\textsf{Isa\kern-0.2exF\kern-0.2exo\kern-0.2exR}%
\xspace}
\newcommand\ceta{\textsf{C\kern-0.2exe\kern-0.5exT\kern-0.5exA}\xspace}
\newcommand\RB{\sh{remote_build}\xspace}

\begin{document}

\maketitle

\begin{abstract}
This is an introduction to the \RB tool for transparent remote session builds.
The intended workflow for a user is to locally issue a build command for some
session heap images and then continue working, while the actual build runs on
a remote machine and the resulting heap images are synchronized incrementally
as soon as they are available.
\end{abstract}

\section{Introduction}

Many Isabelle users are familiar with the following scenario once their formal
developments reach a certain size:
\begin{quote}\sl
You make a change high up in the theory hierarchy but then want to continue
with your latest theory that is a leaf of the hierarchy. So you initiate a
build of an Isabelle session that covers all imports of your current theory
and \ldots take a break, since all the CPU cycles and gigabytes of RAM your
computer has to offer are needed to finish the build within the next few hours
and you will not even be able to read emails on the same machine in the
meantime.
\end{quote}
While the above might be slightly exaggerated, it is not that far from the
truth. For example, building
\isafor/\ceta\footnote{\url{http://cl-informatik.uibk.ac.at/isafor}} 
takes almost four hours on my current machine if only a single process is
available, and requires at least 16\,GB of RAM to succeed at all.

This is in contrast to only around one hour build time on a workstation with 12
processes available. Thus the obvious solution is that you build session heap
images not on your local machine but instead on some decent remote machine.
Often, you will still want to continue work on your local machine once the
build is finished. So you have to copy the remotely built heap images to your
local machine in such a way that Isabelle does not initiate a new build.

It is surely possible to do all this by hand, but it tends to get tedious after
the $n$th repetition. Which is why
I introduce the Isabelle add-on tool \RB, enabling transparent remote session
builds. The
intended workflow for a user is to locally issue a build command for some
session heap images and then immediately continue work without the performance loss that
often comes with time and computation intensive Isabelle builds.
Meanwhile, the actual build is started on another machine and the resulting heap
images are synchronized incrementally as soon as they are available.

\section{Invoking the Build Process}

The \RB tool is implemented in
Isabelle/Scala\footnote{\href{http://isabelle.in.tum.de/doc/system.pdf\#page=32}{\nolinkurl{http://isabelle.in.tum.de/doc/system.pdf}}
(Chapter 4)} and comes with a command line
interface. Its usage is:
\begin{bash}
Usage: isabelle remote_build [OPTIONS] SESSIONS ...

  Options are:
    -B DIR       base directory for remote Isabelle installations (default:
                 $REMOTE_BUILD_REMOTE_BASE, or if former not set ~)
    -d DIR       include session directory
    -r HOST      remote host name (default: $REMOTE_BUILD_REMOTE_HOST)
    -o OPTION    add option for remote isabelle call, e.g., -o -d -o '$ISAFOR'
    -i           incremental: only synchronize heap images that are newly
                 built on the remote host (default: synchronize all session
                 heaps together with their ancestors)
    -P PROXY     connect to remote host via proxy jump; PROXY may either be a
                 HOST or a specification HOST:PORT (default PORT: 2222)
    -v           be verbose

  Build and copy heap images, observing implicit settings:

  REMOTE_BUILD_REMOTE_HOST="..."
  REMOTE_BUILD_REMOTE_BASE="..."
\end{bash}
\lstset{xleftmargin=1em}%

In order for \RB to work properly, we need (at least) two computers, a local
machine $L$ and a remote machine $R$, with Isabelle installed. The respective
installations should be reasonably similar (meaning if one of them is
\nolinkurl{x86_64-linux}, the other should be too; and of course the Isabelle
versions should coincide).
Also, the sessions you want to build and corresponding theory sources have to
be present on both machines (for \isafor, I achieve this for example by using
two clones of its mercurial repository, one on $L$ and one on $R$).
Moreover, communication between $L$ and $R$ runs through SSH and the
\tool{rsync}\footnote{\url{https://rsync.samba.org/}} utility is used for heap
image synchronization.

By default, the Isabelle installation on $R$ is expected to be located in the
user home directory. This can be overwritten by explicitly setting the
remote base directory with \sh{-B}, or made persistent in
\sh{$ISABELLE_HOME_USER/etc/settings} by setting
\sh{REMOTE_BUILD_REMOTE_BASE}.

As for the standard \sh{build} tool of Isabelle, (local) session directories
can be specified via \sh{-d}. Usually, this has to be reflected on the remote
side. A general way of passing options to the Isabelle process invoked on $R$
is by \sh{-o} (which takes a single word, no spaces, as argument).

The hostname/IP address of $R$ can be set explicitly using \sh{-r} or made
persistent by
setting \sh{REMOTE_BUILD_REMOTE_HOST}.

If \sh{-i} is set, then \RB enters incremental mode and only synchronizes heap
images that are generated during the current build. This might occasionally be
useful to save some time (for example, you might already have started to
manually copy heap images from $R$ that existed before the build was
initiated). The default behavior is to synchronize all ancestors of the built
sessions.

If $R$ is not directly available via SSH, a proxy $P$ can be specified using
\sh{-P}, which works as long as $P$ is reachable via SSH from $L$ and $R$ is
reachable via SSH from $P$.

\paragraph{Usage examples.}
This is, for example, how I build the whole of \isafor/\ceta from my office
(``remote host'' and ``remote base'' are implicit in my local settings):
\begin{bash}
isabelle remote_build -d'$ISAFOR' -o-d'$ISAFOR' CeTA
\end{bash}
If I want to do the same from at home, I have to provide a proxy, since the
``build machine'' of our research group is not directly available from the
outside:
\begin{bash}
isabelle remote_build -P proxy.uibk.ac.at -d'$ISAFOR' -o-d'$ISAFOR' CeTA
\end{bash}

\section{Installation Instructions}

The \RB tool is part of the \isafor/\ceta project since version \isaforversion
and compatible with \isaversion.  Its sources reside in
\href{http://cl2-informatik.uibk.ac.at/rewriting/mercurial.cgi/IsaFoR/raw-file/3735ca6ffdb9/src/remote_build.scala}{\file{src/remote_build.scala}}.

Once you obtained the sources, the following steps are required to make \RB
locally available as Isabelle tool. Start by compiling the sources
\begin{bash}
isabelle scalac remote_build.scala
\end{bash}
which should create the two files:
\sh{Remote_Build.class} and \sh{Remote_Build$.class}.
Then, assemble a JAR archive \file{remote_build.jar} via:
\begin{bash}
jar cevf Remote_Build remote_build.jar \
                                Remote_Build.class 'Remote_Build$.class'
\end{bash}
Now, say in a directory \file{tools/}, create the tool wrapper \file{remote_build} with content
\begin{bash}
#!/usr/bin/env bash
$ISABELLE_TOOL scala /path/to/remote_build.jar "$@"
\end{bash}
and register it as Isabelle tool by adding
\begin{bash}
ISABELLE_TOOLS="$ISABELLE_TOOLS:/path/to/tools/"
\end{bash}
to \file{$ISABELLE_HOME_USER/etc/settings}.

\section{Some Further Details and Troubleshooting}

The \RB tool employs the available Isabelle/Scala interface to the
\tool{JSch}\footnote{\url{http://www.jcraft.com/jsch/}} Java implementation of
the SSH2 protocol.
Since the available interface does not cater for password authentication (which would be
cumbersome anyway), the involved SSH connections assume key-based
authentication. However, the current version does not seem to support ECDSA
based host keys.\footnote{The only kind of keys I actually tested is RSA.}
Therefore, it will sometimes be necessary to set up an RSA
host key.

To find out what kind of keys are currently known for a given host \sh{host},
use
\begin{bash}
ssh-keygen -F host
\end{bash}
which looks up host keys in \file{~/.ssh/known_hosts}.
To obtain an RSA key for \sh{host}, use:
\begin{bash}
ssh-keyscan -t rsa host
\end{bash}
Its output can directly be appended to the list of known hosts as follows:
\begin{bash}
ssh-keyscan -t rsa host >> ~/.ssh/known_hosts
\end{bash}

In case a proxy $P$ is used between $L$ and $R$, 
\RB establishes, behind the scenes, the following SSH connections.
First a connection from $L$ to $P$ with port-forwarding from $L:2222$ to the
SSH daemon of $R$. That is, akin to:
\begin{bash}
ssh -L 2222:R:22 P
\end{bash}
And in addition the actual connection between $L$ and $R$ that is carried
inside the above port-forwarding channel. Which you could establish on a
command line via \sh{ssh -p 2222 localhost}.
This setup, causes the peculiarity that an entry for \sh{[localhost]:2222}
is needed in \file{~/.ssh/known_hosts} that provides an RSA key for the remote
host $R$ (so, if you change your remote host, also the key for
\sh{[localhost]:2222} has to change, even though the hostname of the entry did
not).


\end{document}